# Length L-function for Network-Constrained Point Data

Zidong Fang [a, b], Ci Song [a, b], Hua Shu [a, b], Jie Chen [a, b], Tianyu Liu [a, b], Xi Wang [a, b], Xiao Chen [a, b], Tao Pei [a, b, c] *

*a. State Key Laboratory of Resources and Environmental Information System, Institute of Geographic Sciences and Natural Resources Research, Chinese Academy of Sciences, Beijing 100101, China*

*b. University of Chinese Academy of Sciences, Beijing 100049, China*

*c. Jiangsu Center for Collaborative Innovation in Geographical Information Resource Development and Application, Nanjing 210023, China*

[*] Corresponding author: Dr. Tao Pei

*E-Mail address:* peit@lreis.ac.cn (T. Pei)

*Zip code:* 100101

*Phone number:* 01064888960

# Length L-function for Network-Constrained Point Data

*Network-constrained points are referred to as points restricted to road networks, such as taxi pick-up and drop-off locations. A significant pattern of network-constrained points is referred to as an aggregation; e.g., the aggregation of pick-up points may indicate a high taxi demand in a particular area. Although the network K-function using the shortest-path network distance has been proposed to detect point aggregation, its statistical unit is still radius-based. R-neighborhood, in particular, has inconsistent network length owing to the complex configuration of road networks which cause unfair counts and identification errors in networks (e.g., the length of the r-neighborhood located at an intersection is longer than that on straight roads, which may include more points). In this study, we derived the length L-function for network-constrained points to identify the aggregation by designing a novel neighborhood as the statistical unit; the total length of this is consistent throughout the network. Compared to the network K-function, our method can detect a true-to-life aggregation scale, identify the aggregation with higher network density, as well as identify the aggregations that the network K-function cannot. We validated our method using taxi trips' pick-up location data within Zhongguancun Area in Beijing, analyzing differences in maximal aggregation between workdays and weekends to understand taxi demand in th*e *morning and evening peak.*

# 1. Introduction

Location-based phenomena in geography, economics, ecology, and epidemiology can be naturally abstracted into a point. Points can be classified into two types based on space—one is a free point which can be located anywhere within the planar space, and the other is a constrained point which is located only in a restricted space. For example, pedestrian dwell points in an urban road network are constrained points.

Among point patterns, aggregation was the most common and significant. Aggregation can be identified in planar if the average number of points within a circular neighborhood centered on a certain point is statistically greater than expected for a complete spatial random (CSR) distribution (Kiskowski, Hancock, & Kenworthy, 2009). That point is the aggregation center, and the aggregation scale is based on the radius of the neighborhood, which indicates the location and extent of the aggregation, respectively. In contrast to point clusters, which are collections of adjacent points, an aggregation represents the most intense gather in the research area. Detecting and quantifying point aggregations plays an important role in their application to road networks, e.g., detecting spatial aggregations of traffic crashes plays an essential role in the pursuit of improving transit safety and sustainability in urban road networks (Nie, Wang, Du, Ren, & Tian, 2015). Comparing aggregations of origin and destination points of taxis is informative for taxi route selections (Deng et al., 2019); analyzing aggregations of roadside populations with three Acacia species indicated that roadwork activities may have a stronger controlling influence on population dynamics than environmental determinants (Spooner, Lunt, Okabe, & Shiode, 2004). Although identifying and quantifying aggregations is important for understanding point patterns in road networks, very few practical methods in the context of urban networks have been proposed.

The spatial statistic-based method is most effective and precise to identify point aggregation, such as that used by Moran's I (Moran, 1950), Getis-Ord G statistics (Getis & Ord, 1992; Ord & Getis, 1995), and Ripley's K-function (Ripley, 1976, 1977). The main purpose of these methods is to create a new significant test for spatial homogeneity. If the null hypothesis is rejected, a global or local aggregation anomaly is detected. Among existing spatial statistic-based methods, Ripley's K-function is among the most effective since it can detect point patterns at a series of scales. Aware of its powerful capability in detecting aggregation, scholars have extended Ripley's K-function to the network space, making it adaptable for urban applications. Network K-functions and network cross K-functions using shortest-path distance were proposed to analyze the point distribution in networks (Okabe & Yamada, 2001). Furthermore, studies have shown that planar K-functions tend to overestimate the aggregated tendency based on vehicle crash distribution (Lu & Chen, 2007; Yamada & Thill, 2004, 2007). In these studies, network distances based on the shortest path distance, instead of the Euclidean distance, were used as the distance matrix, while linear segments instead of planar circles were utilized as statistical units, bringing the detection results closer to the actual situation.

The shortcomings of the existing network K-functions have led to a few challenges. First, there is unfairness in measuring heterogeneity; the network itself is a clustered subset of the planar region and the radius-based $r-neighborhood$ of points in the network may have different shapes and sizes (see Figure 1). For example, if all points are randomly distributed, the point at the intersections is more likely to be identified as the aggregation, since the road network covered by the $r-neighborhood$ might be longer at the intersection (such as that shown on the right in Figure 1). The intersection will thus possess more points. Secondly, since it is challenging to obtain a consistent size for the $r-neighborhood$ in a network, the expectation value of network K-function under CSR, i.e., benchmarks of these models can only be stemmed from the Monte Carlo simulation (Yamada & Thill, 2007). This makes them incapable of accurately quantifying the scale of the aggregation, as well as wasting a lot of time in computing the benchmark under different road networks. This benchmark, however, is the bridge from the K-function to its advanced version, i.e., the L-function (Besag, 1977; Ehrlich et al., 2004; Kiskowski et al., 2009), which can reveal the center and scale of an aggregation; its local version can identify aggregations based on corresponding scales. In addition, when there are multiple aggregations in a research area, the L-function can focus on the dominant one, i.e., the maximal aggregation, thus helping to isolate the main crux of finding the strongest aggregation (Kiskowski et al., 2009).

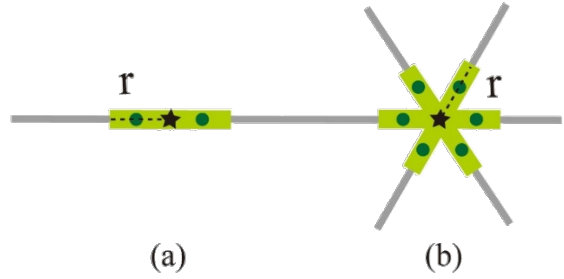

**Figure 1** $r-neighborhood$ in Network K-function. (a) and (b) have the same radius but different lengths in networks: (a) the length the $r-neighborhood$ is $2r$, and (b) the length of the $r-neighborhood$ is $6r$.

In this study, we first propose the $h-neighborhood$, which is more suitable for networks because regardless of where the center point is located, its length remains the same, thus overcoming the weakness of existing definitions of neighborhood in the network K-function. In addition, we devised a method for counting points within the $h-neighborhood$ using the $k^{th}$-Nearest Neighbor distance. Using this, we extended the improved network K-function to L-function after deriving the theoretical value of the K-function under CSR, named the Length L-function. We verified our method using synthetic cases that simulate various types of aggregations in road networks to demonstrate the superiority of the Length L-function in identifying maximal aggregation as compared to the existing network K-function. We validated our method by detecting the aggregation of pick-up points of a taxi trip in Beijing to examine taxi demand in different time periods within the research area.

## 2. Basic concepts

**Definition 1: Road network distance**—This is the shortest path length in the network. Calculating the shortest path between two points on a road map may be modeled as a special case of the shortest path problem in undirected graphs, using Dijkstra's algorithm (Dijkstra, 1959), wherein vertices correspond to intersections and edges correspond to road segments, each weighted by the length of the road segment.

**Definition 2: *h-neighborhood*** —The total network length of a subset of network space centered on a certain point is $h$ (see **Figure 2**). It has two distinctive properties: first that the road network distance from each end point to the center point is the same, and second that the $h-neighborhood$ might have different shapes owing to the configuration of the network(s).

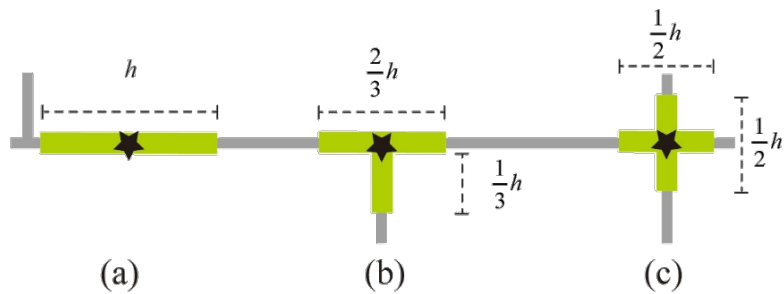

**Figure 2.** The illustration of $h-neighborhood$. (a), (b), and (c) are the $h-neighborhood$ areas centered at different places in road networks: (a) straight roads, (b) forked roads, and (c) crossroads.

The key step in the L-function process is to calculate the number of points within a

$h-neighborhood$ centered on a certain point $p_i$; it is difficult to do so because the shape of $h-neighborhood$ varies with the location of $p_i$ in different networks. Using the $k^{th}$-Nearest Neighbor distance of $p_i$ (denoted as $d_{i,k}$) it is much easier to count target points. For each $d_{i,k}$ (from $k=1$ to $k=n$, $n$ is the number of points in the research area), we calculate the length of the network that the $r-neighborhood$ ($r=d_{i,k}$) occupies, denoted as $L_{i,k}$. Once $L_{i,k+1}$ exceeds $h$, the number of points within the $h-neighborhood$ is $k$ (see **Figure 3**).

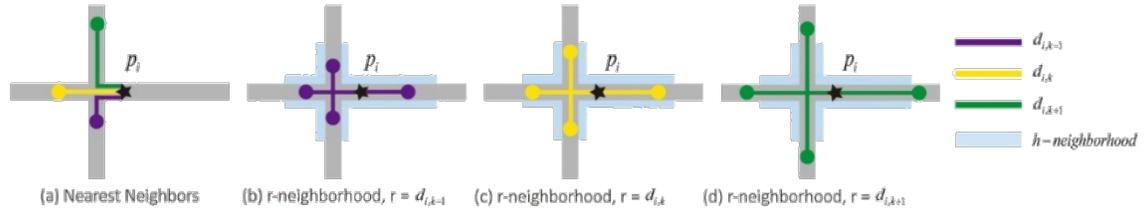

**Figure 3.** Calculating the number of points within an $h-neighborhood$ using the $k^{th}$-Nearest Neighbor distance. (a) Nearest Neighbors and corresponding Nearest Neighbors distance of point $p_i$, (b) $r-neighborhood$ when $r=d_{i,k-1}$, (c) $r-neighborhood$ when $r=d_{i,k}$, (d) $r-neighborhood$ when $r=d_{i,k+1}$. Traverse the $k^{th}$-Nearest Neighbor distance of point $p_i$, namely $d_{i,k}$ ($k=1,...,n$), when the total length of the $r-neighborhood$ with $r=d_{i,k+1}$ exceeds $h$, the number of points within the $h-neighborhood$ (light blue area) is $k$.

**Definition 3: Point process intensity**—This is the expected number of points per unit length of network, denoted by $\lambda$, and is usually estimated by $\hat{\lambda}=n/L_A$; where $n$ is the number of points in the research area and $L_A$ is the total length of the research network. Although it is difficult to precisely identify the boundary of the research area in the real world with a point process, the intensity parameter can be estimated using the second-order properties of points (Pei et al., 2015; Pei, Zhu, Zhou, Li, & Qin, 2007); this is calculated through $d_{i,1}$. The $1^{st}$-Nearest Neighbor distance to each point is calculated as follows:

$$\hat{\lambda} = \frac{1}{E(d_{i,1})} = \frac{n}{\sum_{i=1}^{n} d_{i,1}} \quad (1)$$

**Definition 4: Complete spatial randomness point processes in networks**—During completely spatial random generation, points occur independently and completely randomly within the research network. CSR points can be represented by a homogeneous spatial Poisson process (Dixon, 2001), which is a random counting measurement method (Shu et al., 2020). For a subset $s$ of the research network space, the number of points within $s$, namely $N_s$, follow a Poisson distribution, which is calculated using the following equation:

$$P(N_s = n) = \frac{(\lambda L_s)^n}{n!} e^{-\lambda L_s}, (n = 0,1,2...) \quad (2)$$

where $\lambda$ is the point process intensity in **Definition 3** and $L_s$ is the size of $s$, i.e., the length that $s$ occupies in the network space.

## 3. Length K-function and Length L-function in networks

In this section, we describe the derivation of the length K-function using the

$h-neighborhood$, as well as its theoretical value under CSR. Then, the length L-function is developed based on the length K-function, depicting and amplifying the deviation from the CSR distribution. The local length L-function is also illustrated for a refined understanding of aggregation.

**3.1 Length K-function**

Similar to Ripley's K-function for point-pattern analysis in planar space, the length K-function is defined as the expected number of additional points within the $h-neighborhood$ of an arbitrary point normalized by the point process intensity. The derived formula for $K(h)$ is as follows:

$$K(h) = \lambda^{-1} \frac{\sum_i \sum_j \sigma_{ij}(h)}{n}, (i, j = 1, 2, ..., n; i \neq j) \tag{3}$$

$$\sigma_{ij}(r) = \begin{cases} 1, d_{i,j} \leq h \\ 0, d_{i,j} > h \end{cases} \tag{4}$$

where $h$ is the designated network distance, $n$ is the number of points in the research area, and $\lambda$ is the intensity of the point process.

**3.2 Length L-function and local Length L-function**

In point-pattern analysis, CSR describes a point process whereby point events occur within a given study area in a completely random distribution. CSR is often applied as a standard or benchmark against which datasets are tested. In this study, inferences about

CSR points assist in deriving the length L-function from the length K-function.

Using the aforementioned definition of the length K-function, the expected number of additional points within an $h-neighborhood$ is $\lambda K(h)$. Combining this with the properties of CSR points, we can conclude that $\lambda K(h) = \lambda h$. Thus, the expectation for the length K function with a homogeneous Poisson point process is:

$$E(K(h)) = \frac{\lambda h}{\lambda} = h \tag{5}$$

By normalizing the length K-function with its expected value, we can obtain the length L-function—the expectation is zero at any scale (i.e., distance) for CSR points. In this way, we can amplify and analyze the deviation from the CSR point using the following equation:

$$L(h) = K(h) - h = \lambda^{-1} \frac{\sum_i \sum_j \sigma_{ij}(h)}{n} - h, (i, j = 1, 2, ..., n; i \neq j) \tag{6}$$

The local length L-function for each point is defined as:

$$L_i(h) = K_i(h) - h = \lambda^{-1} \frac{\sum_j \sigma_{ij}(h)}{n} - h, (i, j = 1, 2, ..., n; i \neq j) \tag{7}$$

where $h$, $\lambda$, $n$, and $\sigma_{ij}(h)$ in Equations (5), (6), and (7) have the same meanings as in Equation (3).

## 4. Experiment with synthetic data

This section is aimed at testing the correctness and effectiveness of the length L-function. First, the theoretical length K-function and L-function under CSR, over a range of scales, were tested using Monte Carlo simulations. Following this, we generated synthetic point datasets with different aggregations to test the capabilities of identifying aggregations. At the same time, we describe the detailed steps involved in detecting and extracting the maximal aggregation. We also compared the identification results of the existing network K-function and the length L-function.

**4.1 Monte Carlo tests of null models**

Under the null hypothesis of CSR, the expectation of $K(h)$ is $h$, and the expectation of $L(h)$ is zero. Monte Carlo simulations were used to test our hypothesis, wherein we used three types of datasets to simulate classical road network patterns: a grid pattern with 500 CSR points, a radial pattern with 300 CSR points, and a hybrid pattern with 300 CSR points; each dataset has different point process intensities in the network space. For each road network pattern, 50 simulations were performed to obtain the average value.

Figure 4 shows the results of the CSR points for each row corresponding to a network pattern. The dark blue line and dark purple lines with stars correspond to the theoretical $K(h)$ and $L(h)$ curves; the light blue and light purple lines correspond to the simulated means of $K(h)$ and $L(h)$ curves with 50 runs. The blue and purple belts are the 95% confidence bands based on the Monte Carlo simulations. From these figures, we can conclude that both $K(h)$ and $L(h)$ curves fall between the 95% confidence intervals, and that the simulated means almost match the theoretical curves. The theoretical null models of $K(h)$ and $L(h)$ are thus correct and can be used as benchmarks for determining point aggregation patterns.

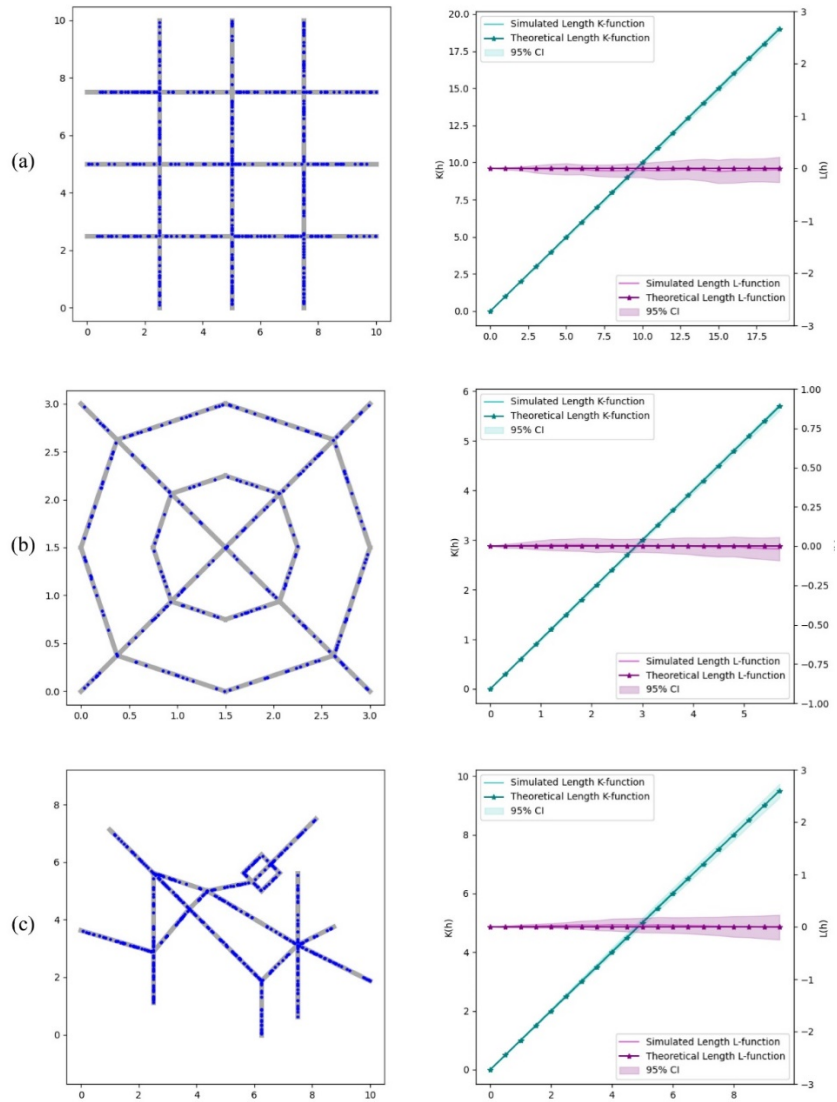

**Figure 4.** Monte Carlo simulation of the length K-function and the length L-function with road networks in various patterns: (a) The grid pattern, (b) the radial pattern, and (c) the hybrid pattern. Each row shows the data sample of that pattern, its theoretical value and the simulated value with 50 runs of the length K-function and the length L-function.

## 4.2 Detecting and extracting aggregation with Length L-functions

We tested our method with simulated road networks in a hybrid pattern and designed four cases of synthetic data to verify the effectiveness of our methodology. Case 1 (see **Figure 5a**) consists of two parts: a linear aggregation with a length of 1.0 and 100 points, and 50 random points on the simulated road network. Case 2 (see **Figure 5b**) consists of two parts: a radial aggregation with a total length of 3.0 and 120 points, which has six equal-length, equal-density branches and 50 random points on the simulated road network. Case 3 (see **Figure 5c**) consists of three parts: the linear aggregation in Case 1, the radial aggregation in Case 2, and 50 random points on the simulated road network. Case 4 (see **Figure 5d**) consists of two parts: the linear aggregation in Case 1 and 300 random points on the simulated road network. In these cases, aggregation scales of the linear aggregation should be $h=1.0$ or $r=0.5$ when detected using the network K-function and the length L-function, respectively. The aggregation scale of the radial aggregation should be $h=3.0$ or $r=0.5$ when detected using the network K-function and the length L-function, respectively. The scale detected using the network K-function corresponds to the radius of aggregation, and the scale detected using the length L-function corresponds to the total length of the road network occupied by the aggregation. Thus, although the values of $\hat{r}$ and $\hat{h}$ may be very different, they indicate the same aggregation scales.

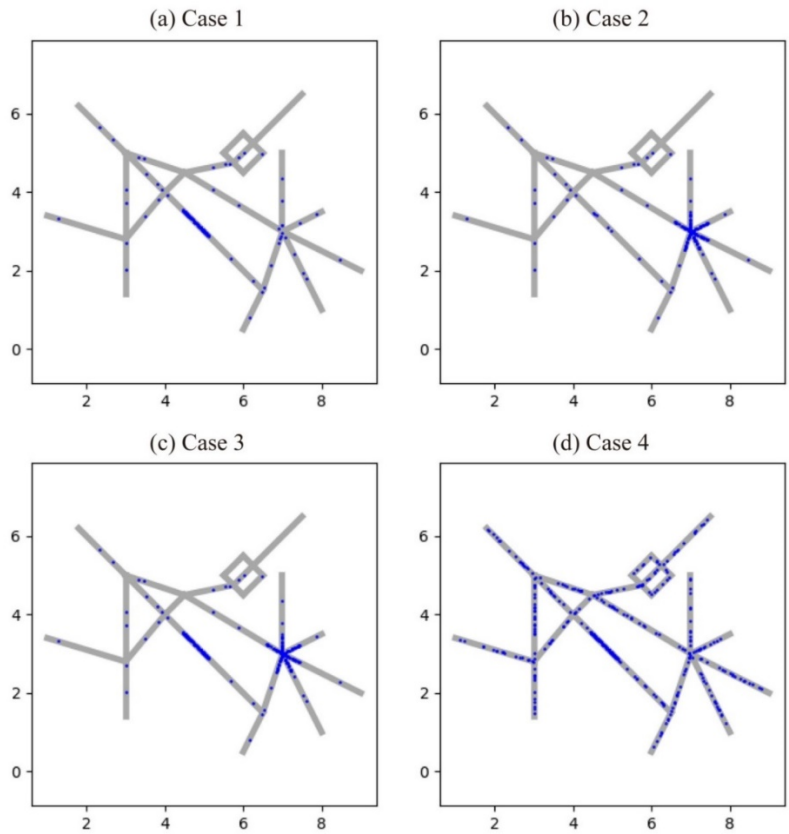

**Figure 5.** Synthetic data with aggregations. (a) Case 1 consists of a linear aggregation and 50 CSR points, (b) Case 2 consists of a radial aggregation and 50 CSR points, (c) Case 3 consists of the linear aggregation in Case 1, the radial aggregation in Case 2, and 50 CSR points (the point process intensity of the linear aggregation is higher for radial aggregation), and (d) Case 4 consists of the linear aggregation in Case 1 and 300 CSR points.

For each case, we detected the maximal aggregation scale with the network K-function and the length L-function, respectively; we then extracted the maximal aggregation with the local version of the corresponding functions. When using the network K-function, we regard the scale (denoted as $\hat{r}$) when the degree of deviation, from the observed K value to the random simulated K value, is the highest as the maximal aggregation scale. If the deviation is not obvious, however, it is considered that there is no aggregation in this dataset (Yamada & Thill, 2007).

The length L-function is an accumulative function, meaning that the effects at larger distances are confounded with those at smaller distances. Using the derivative of the L-function, namely $L'(h)$, can better describe the aggregation scale (Kiskowski et al., 2009). The steps for estimating the maximal aggregation scale using the length L-function are as follows: first, denote the distance that maximizes $L(h)$ as $[L_{max}]$ and denote the distance that minimizes $L'(h)$ as $[L'_{min}]$; second, find the most obvious minimum $[L'_{min}]$ that is greater than $[L_{max}]$ and denote it as $min[L'_{min}]$; and finally, obtain $\hat{h} = min[L'_{min}]/2$ (Shu et al., 2020). In addition, each minimum $L'(h)$ could indicate an aggregation at a certain scale.

After obtaining the scale of the maximal aggregation, $\hat{r}$ and $\hat{h}$, we extracted aggregation with the detected scale and the point possessing the maximal local function value. Taking the local length L-function [see Equation (7)] as an example, the specific extraction steps are as follows: first, calculate the local L-function value of each point with the detected maximal aggregation scale $\hat{h}$ in the dataset, and then select the point with the maximum local L function value, which means that this point is the aggregation center. Finally, points within the $h-neighborhood$ of the aggregation center were merged as the results of extracting maximal aggregation.

Figure 6 shows the results of synthetic data for each row corresponding to the case in Figure 5. Each row lists the observed network K-function and the 10-fold simulation of CSR points with the same quantity as research data; the extracted maximal aggregation with the network K-function, the length L-function and its derivative, and the extracted maximal aggregation with the length L-function.

Figure 6(a) shows four results for Case 1. The aggregation scale is $\hat{r} = 1.0$, as detected using the network K-function and $\hat{h} = 2.5/2 = 1.2$, as detected using the length L-function. For extracting maximal aggregation, the extraction precision was 91.74% and the recall was 100% when using the network K-function; the extraction precision was 95.24%, and the recall was 100% when using the length L-function. In this case, the length L-function has a better ability to detect aggregation scales and extract aggregations than that of the network K-function.

Figure 6(b) shows four results for Case 2. The aggregation scale is $\hat{r} = 1.0$, detected using the network K-function, while $\hat{h} = 5.4/2 = 2.70$ detected using the length L-function (although the scale $h = 4.6$ is the first minimum larger than $[L_{max}]$, it is not obvious enough to indicate an intense aggregation). For extracting maximal aggregation, the extraction precision was 90.23% and the recall was 100% when using the network K-function; the extraction precision was 92.92%, and the recall was 87.5% when using the length L-function. In this case, the network K-function and the length L-function have similar identification abilities, but a more accurate result can be obtained using the length L-function.

Figure 6(c) shows four results for Case 3. The aggregation scale is $\hat{r}=1.0$, as detected using the network K-function, and $\hat{h}=4.6/2=2.3$, as detected using the length L-function. In this case, the designed network intensity of linear aggregation is $I_{Net}=\frac{100}{1.0}=100.0$, which is larger than that of the radial aggregation, which is $I_{Net}=\frac{120}{3.0}=40.0$. Thus, the ideal result of maximal aggregation identification is linear because we should focus on higher road network intensity for constrained points. The expected results were obtained using the length L-function with a precision of 93.46% and recall of 100%. However, when detecting with the network L-function and a scale of $r=0.5$, the corresponding $r-neighborhood$ may contain more points when centered on the radial aggregation than on the linear one. Thus, the network K-function regarded the radial one as the maximal aggregation and failed to identify the ideal aggregation.

Figure 6(d) shows the four results of Case 4. The network K-function has no obvious maximal deviation, which means it fails to detect maximal aggregation while the length L-function can be detected as $\hat{h}=2.8/2=1.4$. For extracting aggregations, the extraction precision with the length L-function was 85.47% and the recall was 100%. In this case, the designed network intensity of linear aggregation is $I_{Net}=\frac{100}{1.0}=100.0$, which is larger than that of all the points, i.e., $I_{AllNet}=\frac{400}{35.97}=11.12$. Owing to the high intensity of points on the entire road network in this case, the $r-neighborhood$ around multi-fork intersections would contain more points than around the straight road, even though there may be an aggregation on the straight road. Such unfairness when counting in the network K-function interferes with the statistical process, and the real aggregation cannot be distinguished from large amounts of background CSR points.

Based on the above cases, it can be found that in both detecting scales and extracting maximal aggregations in the network, the length L-function performs better than the network K-function. Compared to the network K-function, the length L-function can amplify deviation between the observed and theoretical values, paying more attention to the aggregation with higher point process intensities in the network will achieve more equitable statistics. Thus, it could not only detect the point set with higher network intensity as the maximal aggregation, but can also identify the aggregation despite a number of background noise points, which the network K-function cannot achieve. This means that the length L-function is more suitable for detection of aggregation in road network spaces.

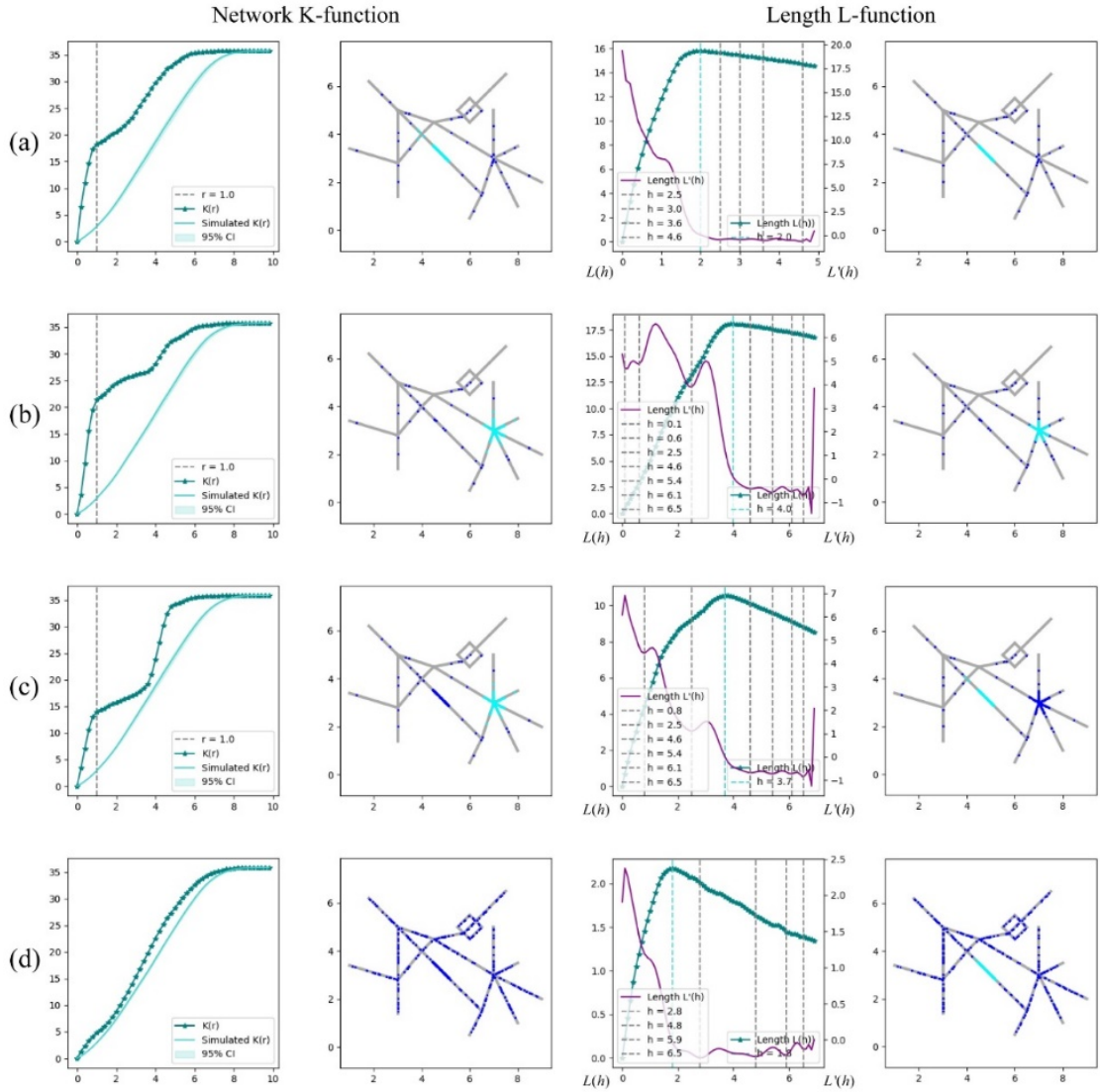

**Figure 6.** Results of the network K-function and the length L-function. From left to right are the results of the network K-function (the scale when the difference between the observed network K value and the simulated theoretical K value is the largest, and is represented by grey dotted lines), the extraction results based on the local network K-function, the length L-function (the maximal value is represented by the blue dotted line) and its derivatives (extreme values are represented by grey dotted lines), and the extracted aggregation with the local length L-function.

## 5. Case study using real-world data

In previous studies on detecting traffic aggregation with either the K-function or the L-function, data is invariably added up over a long period of time, making the point aggregation more obvious. For example, taxi GPS data are accumulated within one week as the experimental data so that aggregation can be detected (Shu et al., 2020). To understand real-time and short-term taxi demand, we detect the aggregation on the road network using the length L-function and distinguish the aggregation from the number of noise points.

The case study used taxi GPS trajectory data on October 21$^{st}$, 2014 (a workday) and October 25$^{th}$, 2014 (a weekend) in Beijing. After matching the data to the road network, we extracted the pick-up locations of each trip and identified the maximal aggregation in the dataset during the morning (7:00–9:00) and evening (17:00–19:00) peaks to identify and understand the subtle changes of taxi demand hotspots in the study area.

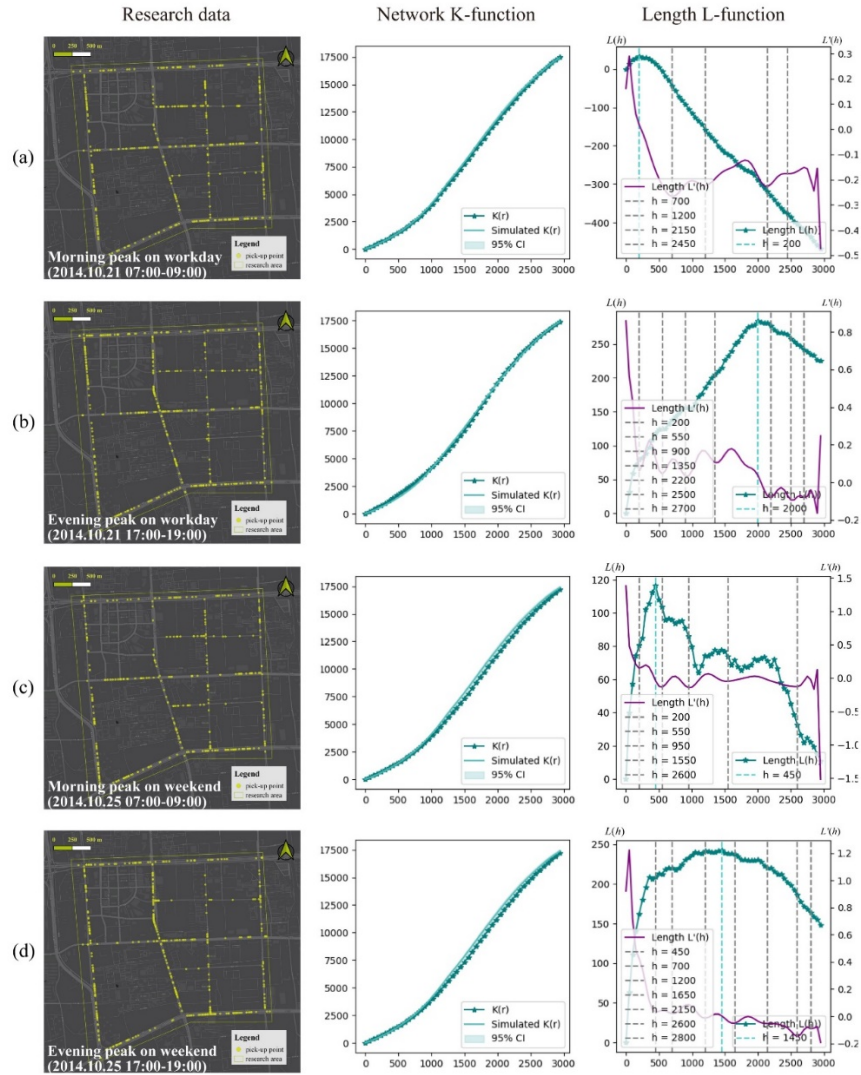

**Figure 7.** Experimental data and its corresponding results determined using the network K-function and the length L-function. From left to right are the experimental point distribution, the results of the network K-function (there is no obvious difference between the observed network K value, represented with dark blue line with star, and the simulated K value of random points colored in light blue), the length L-function (colored in purple with the maximal value is represented by the blue dotted line) and its derivatives (minimum values are represented by grey dotted lines).

The identification results are shown in Figure 7, and signal that the network K-function failed to detect aggregations since there is no obvious deviation from the observed K-function and the simulated K-function (10 simulations of random points with the same number of points as the experimental data). The length L-function worked and detected the corresponding aggregation changes in this area.

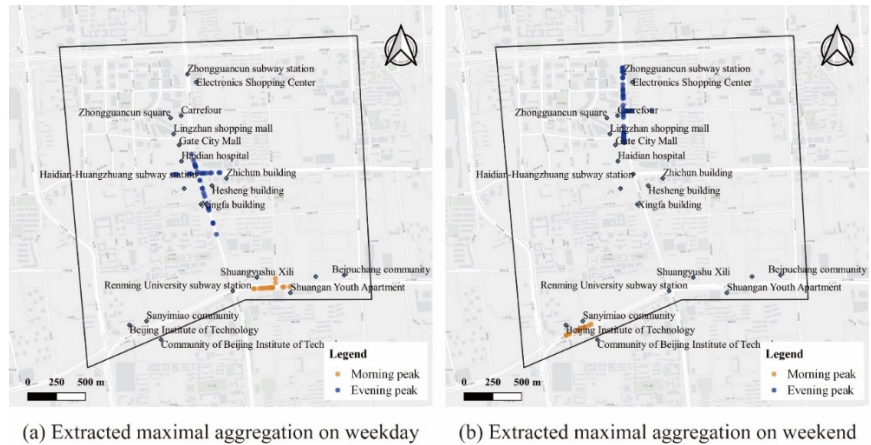

(a) Extracted maximal aggregation on weekday  (b) Extracted maximal aggregation on weekend

**Figure 8.** Extracted maximal aggregation. (a) Aggregation on weekdays and (b) aggregation on weekends, where maximal aggregation during the morning peak is orange and the maximal aggregation during the evening peak is blue.

Maximal aggregation during the morning peak (see the orange points in Figure 8) gathers along residential areas on weekdays or on weekends; however, the aggregation center changed significantly here. It is close to the Renming University subway station and the eastern neighborhood composed of Shuangan Youth Apartment and Shuangyushu Xili, with an aggregation scale of $\hat{h} = 350\ m$; on the weekend, the maximal aggregation is located near the Beijing Institute of Technology and western neighborhood consisting of Sanyimiao community, to the extent of $\hat{h} = 275\ m$. Whether on weekdays or on weekends, the morning peak's maximal aggregations are close to the residential community because residents invariably choose to take a taxi to go out in the morning. It is worth noting that the maximal aggregation is closer to subway station exits on weekdays, because workers usually use a combination of subways and taxis to commute for an optimized transportation experience (Kim, 2018). After a subway trip, residents often choose to take a taxi to go to work. Over the weekend, the detected aggregation is near a residential area in the west, which is not as large as that in the east, but is close to the Beijing Institute of Technology. This indicates that, compared to ordinary residential areas, residents near the university (probably a younger demographic), will travel more in the morning on weekends.

For the aggregation during the evening peak (see the blue points in Figure 8), a center was observed at the Haidian-Huangzhuang subway station on workdays, while its center moved slightly north, which is closer to the several shopping centers on weekends. The aggregation scale is $\hat{h} = 1100\ m$ on weekdays and $\hat{h} = 825\ m$ on weekends, which has almost no change. The reasons that the maximal aggregation during the evening peak is centered at the Haidian-Huangzhuang subway station on weekdays are: (1) The subway station is surrounded by middle schools, office buildings, and hospitals (see **Figure 8**), which are all busy on weekdays, but few people visit them on weekends; thus, there is a higher taxi demand on weekdays than on weekends; (2) As mentioned before, there are many subway exits and office workers prefer to take a taxi here to go home after a subway trip for faster and more convenient commuting. The aggregation center moves slightly northward on weekends because there are several popular shopping malls in the north, such as Carrefour, the Lingzhan shopping mall, and the Electronics shopping center. Residents frequent shopping malls and other entertainment places more than on weekdays (Su, Spierings, Dijst, & Tong, 2020), thus, the demand for taxis near shopping malls and entertainment spaces increases on weekends, while the demand near subway exits' serving commuting workers will decrease.

The results of the length L-function indicate that, for taxi pick-up points in this research area, the maximal aggregation in the morning peak will occur near residential areas while the maximal aggregation in the evening peak will occur in commercial areas; aggregations on weekdays and weekends will change slightly because of the characteristics of the functional area. Overall, the length L-function can help us find subtle changes of aggregation in networks within a short period of time, even when large numbers of background points exist on roads; this could not be achieved by using the pre-existing road network K function. This helps us analyze the travel behavior of residents on a small scale.

## 6. Conclusion

To precisely identify the point maximal aggregation in various networks, we derived the L-function in the 1-D network space, called the Length L-function. This study both alters the distance matrix (from Euclidean distances to network distances in the existing Planar L-function) and presents a novel calculation unit called the $h-neighborhood$ in order to improve the existing network K-function. Wherever the center point is, e.g., at a multi-intersection or on a straight road, the network space size of the $h-neighborhood$ is always $h$, thus overcoming the unfair statistics in the network K-function. This tackles the problem that the expectation value of the network K-function cannot be deduced, thus helping to obtain the L-function.

The length L-function detects aggregation over a range of scales and determines the maximal aggregation in the dataset, which indicates the most intense gathering of points in a given network. Unlike the existing network K-function, it focuses on detecting point aggregations with higher density in road networks; aggregations can be detected even when there are large numbers of noise points in the network, which fits the demand for practical applications in cities.

The case study assessing the Beijing taxi's pick-up point data shows the applicability of our method. Our experimental results were compared to those of the network K-function, indicating that the network K-function cannot be identified when the aggregation is not obvious, or if there are many noise points; our length L-function succeeds regardless of these constraints. Overall, since the length L-function pays more attention to the road network density of the point set, it is more suitable for aggregation detection in the road network space, and is thus more adaptable for research on urban environments.

Future research in this area could include several aspects. First, a study could extend other spatial statistical methods designed for planar space to network space with the aid of the $h-neighborhood$. Second, a future study could focus on both the point aggregation in road networks and on studying point aggregation patterns in other types of networks, e.g., flight networks and social networks.

Table 1 The parameters of aggregation in four cases

| Case | Radial aggregation | | | | Linear aggregation | | | | CSR points |
|---|---|---|---|---|---|---|---|---|---|
| | r | h | Number | NI | r | h | Number | NI | Numbers |
| 1 | / | / | / | / | 0.5 | 1.0 | 100 | 100.00 | 50 |
| 2 | 0.5 | 3.0 | 120 | 40.00 | / | / | / | / | 50 |
| 3 | 0.5 | 3.0 | 120 | 40.00 | 0.5 | 1.0 | 100 | 100.00 | 50 |
| 4 | / | / | / | / | 0.5 | 1.0 | 100 | 100.00 | 300 |

* NI refers to the point process intensity in network space.